\documentstyle[12pt]{article} \begin{document}
\voffset = -2cm
\textheight = 23cm
\begin{center}

{\bf ANOXIA DURING THE LATE PERMIAN BINARY MASS EXTINCITON AND DARK
     MATTER} \\
\vskip 10mm   
     {\bf Samar Abbas }, \\
     {\it Physics Department, Utkal University, \\
     Bhubaneshwar-751004, Orissa, India \\
     e-mail: abbas@beta.iopb.stpbh.soft.net }\\
\vskip 2mm
     {\bf Afsar Abbas }, \\
     {\it Institute of Physics, Bhubaneswar-751005, Orissa, \\
     India \\
     e-mail: afsar@beta.iopb.stpbh.soft.net } \\
\vskip 2mm
     {\bf Shukadev Mohanty }, \\
     {\it Physics Department, Utkal University, \\
     Bhubaneshwar-751004, Orissa, India } \\
\vskip 20mm 
ABSTRACT
\end{center}     
\vskip 10mm
\sf{
                                      
   Recent evidence quite convincingly indicates that the Late Permian
   biotic crisis was in fact a binary extinction with a distinct
   end-Guadalupian extinction pulse preceding the major terminal
   end-Permian Tartarian event by 5 million years. In addition anoxia
   appears to be closely associated with each of these end-Paleozoic
   binary extinctions. Most leading models cannot explain both anoxia and
   the binary characteristic of this crisis. In this paper we show that
   the recently proposed volcanogenic dark matter scenario succeeds in
   doing this. 

\begin{center}
INTRODUCTION 
\end{center}
                                      
   Recently, Knoll, Bambach, Canfield and Grotzingen (Knoll et al. 1996a)
   have suggested a new model wherein the overturn of anoxic deep oceans
   led to the end-Permian Tartarian extinction by the introduction of
   carbon dioxide into surficial environments. This model could explain
   the selectivity of the extinction, with organisms tolerant of elevated
   carbon dioxide levels exhibiting higher degrees of survival across the
   P/T boundary. The C-isotope record also indicates that another anoxic
   event occurred at the end of the Guadalupian, approximately 5 million
   years before the Tartarian extinction.
   
   Knoll et al state that it is possible that the Siberian flood basalt
   volcanic episode (Campbell et al. 1992) could have led to the
   Tartarian overturn by means of tectonic realignment (Knoll et al.
   1996a) . It has also been realized that this extinction was in fact a
   double extinction (Stanley and Yang 1994) , and the work of Knoll et
   al. is consistent with this fact. This double extinction shall be
   referred to as a "binary extinction" in this work; and the anoxia
   appear to be related to the extinctions. In this paper we set forth a
   scenario based on the recently proposed idea of volcanogenic dark
   matter ( Abbas and Abbas 1998 ). This can consistently explain the
   binary nature of the extinction and the associated anoxia, and
   predicts binary anoxia and double extinctions at other major mass
   extinctions as well. \\

\begin{center}   
VOLCANOGENIC DARK MATTER 
\end{center}
                                      
   Dark matter may constitute more than 90 
   and ample evidence in favour of its existence occurs in the form of
   galactic rotation curves, the stability of galactic clusters etc.
   Several candidates have been proposed ( Berezinsky 1993 , Watson 1997
   ). It is probable that dark matter occurs in a clumped form, with
   high-density clumps of dark matter existing within a uniform halo
   background. During the occasional passage of such a clump through the
   Earth dark matter would accumulate in the core and annihilate,
   producing vast quantities of heat (Kanipe 1997) . Abbas and Abbas
   estimate that the heat output can exceed present-day terrestrial heat
   production by five orders of magnitude (Abbas and Abbas 1998) . These
   large quantities of heat will in all likelihood lead to the creation
   of a superplume that initiates, upon arrival at the surface, the
   Siberian flood basalt volcanic episode (Abbas and Abbas, 1998) . This
   volcanism may lead to changes in oceanic circulation patterns by
   tectonic realignment or the creation of new oceanic plumes above
   submarine eruption sites. Such a change could lead to anoxia with the
   consequent terminal P/T mass extinction as envisaged by Knoll et al.
   (Knoll et al. 1996a) . In addition Vermeij and Dorritie (Vermeij and
   Dorritie 1996) pointed out that it is possible that Siberian volcanism
   may have released vast quantities of methane from permafrost and
   continental shelves, which, on oxidation, would have yielded carbon
   dioxide, drawing down oxygen in the process and leading to anoxia.
   
   This volcano-induced extinction would occur after a time interval
   representing the duration between creation of the superplume at the
   core/mantle boundary and arrival of this plume at the surface. With a
   migration rate of a few cm per year, this should be approximately 5
   million years, and this is in fact the interval separating the
   Guadalupian and Tartarian extinctions (Stanley and Yang 1994) .
   
   According to Isozaki (Isozaki 1997a) , results from Japanese and
   British Columbian deep sea cherts indicate that the onset of anoxia
   marked the Guadalupian extinction, and the climax of the anoxia, or
   the superanoxia, coincided with the Tartarian crisis ( Isozaki 1997b ,
   Retallack and Holser 1997 ). This is consistent with the volcanogenic
   dark matter model; the anoxia would persist instead of disappearing
   after some time. The duration of anoxia in this picture is dependant
   on how rapidly ocean circulation patterns can re-oxygenate the seas.
   In the model outlined herein Siberian volcanism may have released vast
   amounts of methane from permafrost which on oxidation would have
   consumed oxygen and led to anoxia; or oxidation of organisms that died
   as a result of dust, blockage of sunlight, noxious gases (eg. sulphur
   dioxide, nitrogen oxides) etc. would have drawn down oxygen, thereby
   leading to anoxia. Further palaeontological work is required to
   clarify whether anoxia is the cause or effect of extincitons.\\

\begin{center}   
CARCINOGENESIS DUE TO DARK MATTER 
\end{center}
                                      
   The direct passage of a dark matter clump itself may lead to the first
   extinction step by causing lethal carcinogenesis in organisms. Zioutas
   ( Zioutas 1990 ) studied the effect of dark matter on living
   organisms, and concluded that dark matter may be responsible for
   mutation and cancers in living beings. Changes of biorhythms depending
   on the direction of flight have been recorded for humans as well as
   fungi during flights across different time zones, and these may be due
   to dark matter. Background radiation can only explain 1 in 20000 of
   the observed spontaneous mutations in Drosophilia; the remainder may
   be due to dark matter interactions. Subsequently Collar (Collar 1996)
   analyzed the effect that highly clumped dark matter may have on the
   biosphere. He discovered that such an event could be highly
   detrimental to life on Earth. The dosage imparted to organisms during
   the passage of a clump core would in principle be roughly comparable
   to the neutron radiaton from a close nuclear explosion protracted over
   a time required for clump core passage. This dose protraction would
   further aggravate these effects. Thus the passage of a clump core
   would induce a large dose of highly mutagenic radiation in all living
   tissue. Collar then proposed that dark matter could have caused
   palaeontological mass extinctions.
   
   The dark matter, being weakly interacting, does not decrease in
   intensity during passage through the oceans and hence immediate
   extinctions on land and sea would result from the clump passage
   through the Earth. The oxidation of the resulting organic matter would
   deplete oxygen from the oceans and atmosphere, while simultaneously
   increasing the levels of carbon dioxide. Normally, as organic material
   descends through the ocean, it is oxidised on the way down. However,
   the full-scale destruction of life would lead to even the surface
   waters becoming anoxic ( Smith 1989 ). Thus the first pulse of anoxia
   would have been a consequence of extinctions. The introduction of
   anaerobic life may further toxify the oceans as some of these
   organisms produce toxins like hydrogen sulphide, and further
   extinctions of those species that survived the clump passage would
   take place. In addition, for low dosages of dark matter the mutations
   engendered would take several generations to cause fatalities. Hence
   this first extinction need not necessarily be a sharp peak; it may
   very well be extended and drawn out.
   
   Thus the passage of the Earth through a clump of dark matter would be
   highly detrimental to life. This leads to mass extinctions of
   phytoplankton and/or larger marine animals, and/or other types of
   marine biota. Their destruction gradually depletes dissolved oxygen at
   a faster rate than can be replaced by dissolving from the atmosphere,
   thereby removing the oxygen even from the surface waters. Thus,
   anaerobic conditions set in and microbes that can respire
   anaerobically thrive. Anaerobic sulphur bacteria produce hydrogen
   sulphide, which is toxic to most marine life (Smith 1989) . \\

\begin{center}   
DISCUSSION AND CONCLUSION 
\end{center}
                                      
   Thus the model described above makes definite predictions, one of
   which is that most major extinctions should be binary at higher
   resolutions. In fact several appear to be so. Stanley and Yang
   (Stanley and Yang 1994) found that the end-Permian extinction, at
   which probably no major asteroidal/cometary impact occurred, was in
   fact binary, with the Guadalupian extinction eliminating 71 
   species, and 80 
   analysis of life across the K/T boundary by MacLeod et al. (MacLeod et
   al. 1997) rules out a single terminal catastrophe and we feel that
   overlapping binary extinction peaks in the framework outlined above
   are consistent with these results. The late Miocene extinction has
   recently been found to be binary ( Petuch 1995 ), with a five million
   year interval between them. The scenario depicted above may be
   applicable to this case also. In fact the volcanogenic dark matter
   scenario ( Abbas and Abbas 1998 ) provides a natural explanation for
   double extinctions ( Abbas, Abbas and Mohanty 1998 ). Here in this
   paper we have shown how, in addition, anoxia can be explained.
   
   Hence several features of mass extinctions have been considered. If,
   as it seems likely, anoxia, periodicity and a binary feature are
   characteristic of the major extinctions of palaeontology, then the
   volcanogenic dark matter scenario emerges as a viable model
   explanation. \\

\newpage   
\noindent {\bf REFERENCES } \\
                                      
\noindent Abbas, S. and Abbas, A. 1998. \\
   {\it Volcanogenic dark matter and mass extinctions.} \\
   {\tt Astroparticle Physics 8(4/12):317-320}  \\
   http://xxx.lanl.gov/abs/astro-ph/9612214  \\
   
\noindent Abbas, S. Abbas, A. and Mohanty, S. 1998.  \\
   {\it Double Mass Extinctions and the Volcanogenic Dark Matter
   Scenario.} \\
   http://xxx.lanl.gov/abs/astro-ph/9805142  \\
   
\noindent Berezinsky, V.S. 1993.  \\
   {\it High Energy Neutrinos from Big Bang Particles.}  \\
   {\tt Nuclear Physics B (Proceedings Supplement) 31:413-427 } \\
   
\noindent Campbell, I.H., et al 1992.  \\
   {\it Synchronism of the Siberian Traps and the Permian-Triassic
   Boundary.} \\
   {\tt Science 258:1760-63 } \\

\noindent Collar, J.I, 1996. \\
   {\it Clumpy Cold Dark Matter and biological extinctions.} \\
   {\tt Physics Letters B 368:266-9 } \\

\noindent Isozaki, Y. 1997a. \\
   {\it Permo-Triassic Boundary Superanoxia and Stratified Superocean:
   Records from Lost Deep Sea.}  \\
   {\tt Science 276:235-238 } \\

\noindent Isozaki, Y. 1997b. \\
   {\it Response to: Timing of Permian-Triassic Anoxia.} \\
   {\tt Science 277:1745 } \\

\noindent Kanipe, J. 1997.  \\
   {\it Dark mater blamed for mass extinctions on Earth.} \\
   {\tt New Scientist Jan. 11 1997, p.14 } \\

\noindent Knoll, A.H., Bambach, R.K., Canfield, D.E. 
and Grotzinger, J.P. 1996a \\
   {\it Comparative Earth History and Mass Extinction.} \\
   {\tt Science 273:452-457 } \\

\noindent Knoll, A.H., Bambach, R.K., Canfield, D.E. 
and Grotzinger, J.P. 1996b \\
   {\it Late Permian Extinctions: Response to Letters.} \\
   {\tt Science 274:1550 } \\

\newpage

\noindent MacLeod, N. et al. 1997 \\
   {\it The Cretaceous-Tertiary biotic transition.} \\
   {\tt J.of the Geol.Soc.of London 154:265-292 } \\

\noindent Petuch, E.J., 1995 \\
   {\it Molluscan diversity of the Late Neogene of Florida. Evidence for a
   Two-Staged Mass Extinction.} \\
   {\tt Science 270:275-7 } \\

\noindent Retallack, G.J. and Holser, W.T., 1997. \\
   {\it Timing of Permian-Triassic Anoxia (Letter)}. \\
   {\tt Science 277:1745 } \\

\noindent Smith, D.G. (ed.-in-chief) 1989. \\
   {\it The Cambridge Encyclopedia of Earth Sciences}, p. 345. \\
   Cambridge University Press, Cambridge. \\

\noindent Stanley, S.M., and Yang, X., 1994. \\
   {\it A Double Mass Extinction at the End of the Paleozoic Era.} \\
   {\tt Science 266:1340-44 } \\

\noindent Vermeij, G.J. and Dorritie, D. 1996. \\
   {\it Late Permian Extinctions - Letter.} \\
   {\tt Science 274:1550 } \\

\noindent Watson, A. 1997. \\
   {\it To catch a WIMP.} \\
   {\tt Science 275:1736-8 } \\

\noindent Zioutas, K., 1990. \\
   {\it Evidence for Dark Matter from Biological Observations.} \\
   {\tt Physics Letters B 242:257-264 } \\
}

\end{document}